\documentclass[a4paper]{article}
\usepackage[affil-it]{authblk}
\usepackage{natbib}
\usepackage{amstext}
\usepackage[margin=1in]{geometry}
\usepackage{amssymb,amsmath}
\usepackage{color}
\definecolor{darkgreen}{rgb}{0,.5,0}
\usepackage{hyperref}

\hypersetup{
  colorlinks=true,
  urlcolor=blue,
  linkcolor=red,
  citecolor=blue
}

\newcommand{\om}{{\omega}}
\newcommand{\bvpi}{{\mbox{\boldmath $\varpi$}}}

\newcommand{\br}{{\bf r}}
\newcommand{\bv}{{\bf v}}
\newcommand{\bn}{{\bf n}}
\newcommand{\bOm}{{\bf \Omega}}
\newcommand{\bJ}{{\bf J}}
\newcommand{\bw}{{\bf w}}
\renewcommand{\i}{\mathrm{i}}
\newcommand{\ee}{\end{equation}}
\newcommand{\be}{\begin{equation}}
\newcommand{\p}{\partial}
\newcommand{\ce}{{\cal E}}
\newcommand{\case}{\textstyle\frac}

\defcitealias{BT2008}{BT}

\begin{document}

\title{\textsc{Scale-invariant Mode \\in Collisionless Spherical Stellar Systems}}

\author{Evgeny~V.~Polyachenko%
  \thanks{Electronic address: \texttt{epolyach@inasan.ru}}}
\affil{Institute of Astronomy RAS, 48 Pyatnitskaya st.,\\119017 Moscow, Russia}

\author{Ilia~G.~Shukhman%
  \thanks{Electronic address: \texttt{shukhman@iszf.irk.ru}}}
\affil{Institute of Solar-Terrestrial Physics RAS, Siberian Branch,\\P.O. Box 291, Irkutsk 664033, Russia}

\date{Dated: \today}
\maketitle

\maketitle

\begin{abstract}
An analytical solution of the perturbed equations is obtained, which exists in all ergodic models of collisionless spherical stellar systems with a single length parameter. This solution corresponds to variations of this parameter, i.e., stretching or shrinking the sphere while preserving the total mass. The system remains in an equilibrium state. The simplicity of the solution allows for explicit expressions for the distribution function, potential, and density at all orders of perturbation theory. This, in turn, helps to clarify the concept of perturbation energy, which, being a second-order quantity in amplitude, cannot be calculated in linear theory. It is shown that the correct expression for perturbation energy, constructed taking into account 2nd order perturbations, and the well-known expression for perturbation energy constructed as bilinear form obtained within linear theory from 1st order perturbations do not coincide. However, both of these energies are integrals of motion and differ only by a constant. The obtained solution can be used to control the correctness of codes and the accuracy of calculations in the numerical study of collisionless stellar models.

{\bf Keywords:} stellar systems, star clusters and associations, stellar dynamics

\end{abstract}

\section{Introduction}

One of the traditional methods for studying the dynamics of perturbations in equilibrium models of spherical stellar systems is the investigation of the evolution of small perturbations. Typically, the main question is whether the equilibrium state described by the distribution function (DF) of stars \(F(\mathbf{r},\mathbf{v})\) and gravitational potential \(\Phi_0(\mathbf{r})\) is stable or unstable.

Along with general stability criteria based on specific theorems~\citep[see, e.g.,][hereinafter BT]{BT2008}, there exists a method for solving the linearized eigenvalue problem. For this, assuming that the perturbations of the gravitational potential \(\Phi(\mathbf{r},t)\) and DF \(f(\mathbf{r},\mathbf{v},t)\) are small and proportional to \(\exp(-\i\omega t)\), one finds the eigenvalues \(\omega\) of the linearized system of equations, consisting of the collisionless Boltzmann equation and the Poisson's equation. The presence of eigenvalues with \({\rm Im}(\omega)>0\) indicates system instability.

Finding the eigenvalues \(\omega\) is a rather laborious task. Except for a few models where the equilibrium potential is harmonic~\citep[see, for example][]{MFE70, PS73, PS74, MPS74}, it is solved using so-called matrix methods. Here, the problem is reduced to numerically finding the roots \(\omega\) of a certain determinant, \({\cal D}(\omega)\equiv \det \|D^{\alpha\beta}(\omega)\|=0\), \(\alpha, \beta=1,2,3...\). For disk models, the matrix method was first proposed by~\cite{Kalnajs_76}, and for spherical systems by~\cite{PS81}. It involves expanding the amplitudes \({\hat\Phi}(\mathbf{r})\) and \({\hat\rho}(\mathbf{r})\) of perturbed potential and density \(\Phi(\mathbf{r},t)={\hat\Phi}(\mathbf{r})\,e^{-\i\omega t}\) and \(\rho(\mathbf{r},t)={\hat\rho}(\mathbf{r})\,e^{-\i\omega t}\) in terms of a so-called biorthonormal set of basic potential-density pairs, \(\Phi^\alpha(r)\) and \(\rho^\alpha(r)\), and obtaining a system of linear equations for the expansion coefficients \(C^\alpha\). Setting the determinant of this system to zero leads to the desired dispersion relation.

This method is applicable to systems with the integrable Hamiltonian \(H_0\), i.e., for equilibrium stellar systems whose potential \(\Phi_0(\mathbf{r})\) allows a transition from coordinate-velocity variables \((\mathbf{r},\mathbf{v})\) to action-angle variables \((\mathbf{J},\mathbf{w})\). In the alternative matrix method proposed by E. Polyachenko~\citep[see][]{Pol_2004, Pol_2005}, the original system of linearized equations is reduced to a standard linear eigenvalue problem in the form \(\omega f_n(\mathbf{J}) = \sum\limits_{n'} \int d\mathbf{J'}\, K_{nn'}(\mathbf{J}, \mathbf{J'})\, f_{n'}(\mathbf{J'})\), where \(f_n(\mathbf{J})\) are the Fourier harmonics of the perturbed DF with respect to the angular variables \(\mathbf{w}\), and \(K_{nn'}(\mathbf{J},\mathbf{J'})\) is the kernel.

Recently, a series of studies has emerged~\citep[][]{HH20, Ham_21, LB2021a, LB2021b} examining the dynamics of perturbations not for the purpose of investigating the stability of equilibrium models, as was the case in previous decades~\citep[see, e.g.,][BT]{FP84, Pal_94, Bertin2014}, but rather to study density and potential fluctuations around equilibrium and their influence on slow relaxation processes, as well as their role in N-body simulations of stellar systems. Stable equilibrium models are considered, and perturbations in them can be caused either by noise associated with a finite number of particles \(N\)~\citep{LB2021a}, or by an external source. In this case, weakly damped oscillations are of interest, which can last for many characteristic crossing times, practically indistinguishable from true neutral eigenmodes~\citep{Wei_94, HBV_20}.

In the case of stable spherical equilibrium systems, there are no discrete modes with \({\rm Im}(\omega)>0\). Due to the time-reversibility of the collisionless Boltzmann equation, there are also no decaying discrete modes with \({\rm Im}(\omega)<0\). The presence of discrete neutral modes, \({\rm Im}(\omega)=0\), is possible only in rare situations. This is related to the presence of resonances between the perturbation waves and the orbital motion of stars, $\omega-{\bf{n}}\cdot{\bf{\Omega}}=0$, where $\bf{\Omega}(\bf{J})=(\Omega_1,\Omega_2,\Omega_3)$ are the frequencies of orbital motion, and ${\bf n}=(n_1,n_2,n_3)$ are integers. Therefore, neutral discrete modes are only possible in the presence of `gaps' in phase space that are free from resonance.\footnote{For radial perturbations, \cite{Mat_90} provided arguments in favor of the possibility of such neutral modes in principle, but did not give specific examples of corresponding DFs.}
It turns out that for such equilibrium models, the complete system of eigenmodes is exclusively represented by the continuous spectrum of van Kampen modes with real frequency \(\omega\) \citep{vK_55}. It should be noted that a perturbation decaying exponentially according to the so-called Landau damping, in which the frequency \(\omega_{\rm L}\) has a negative imaginary part, \(\omega_{\rm L}={\rm Re} (\omega_{\rm L})+\i\, {\rm Im} (\omega_{\rm L})\), \({\rm Im} (\omega_{\rm L})<0\), is not a true decaying eigenmode, but represents a continuous superposition of singular van Kampen modes. To distinguish a perturbation decaying according to Landau from a true eigenmode, we will refer to it as a \textit{quasi-mode}. More detailed dynamics of initial perturbations, represented as a superposition of van Kampen modes, and its connection with Landau quasi-modes for infinite homogeneous gravitating systems, has been traced in~\cite{PSB}, and for the case of shear flows of fluid in~\cite{PS_22}.

For stable systems, the presence of weakly damped Landau quasi-modes plays a crucial role. Their existence allows oscillations, excited, for example, by the close passage of an external perturber, to persist for a long time without damping \citep[see e.g.][]{Wei_94}. However, not every stable system possesses even a single Landau quasi-mode. In terms of van Kampen waves, the existence of a weakly damped Landau quasi-mode with a real part of the frequency \({\rm Re}(\omega_{\rm L})\) means that the amplitudes of van Kampen waves with frequencies \(\omega\) in the vicinity of \({\rm Re}(\omega_{\rm L})\) become particularly large. The absence of such peaks in the frequency spectrum of van Kampen modes implies that perturbations, avoiding the stage of slow Landau exponential decay, almost immediately transits to algebraic decay regime, \(\sim t^{-n}\), \(n>0\)~\citep[see, for example,][]{PS_22, BOY11}. In particular, we have shown (unpublished) that in the case of shear flow in a channel, \(U(y)=y+0.1\,y^3\), which has an inflection point  but is stable according to the \cite{Fj1950} theorem, there are no Landau quasi-modes at all.

Therefore, the search for quasi-Landau modes in stable systems is of interest but tricky. The dispersion equation obtained by any of the matrix methods described above \citep{PS81, Pol_2004, Pol_2005}, is valid only in the upper half-plane of the complex variable \(\omega\), while the frequencies of Landau quasi-modes lie in the lower half-plane of \(\omega\). This fact is related to the principle of causality and has been repeatedly described in the literature, starting with Landau's pioneering work \cite{Lan_46} \citepalias[see also][]{BT2008}. In order to use the dispersion equation \({\cal D}(\omega)=0\) to find the frequencies of Landau quasi-modes, it is necessary to perform an analytical continuation of the function \({\cal D}(\omega)\) into the lower half-plane of the complex variable \(\omega\). \cite{Lan_46} first carried out this procedure for a homogeneous electron plasma. To do this, he deformed the integration contour over the velocity variable \(v\) (only one in his problem), shifting it down into the complex \(v\)-plane so that it passed below all possible resonance points \(v_c\equiv\omega_{\rm L}/k\). This procedure is called the Landau-Lin bypass rule, since~\cite{Lin55} derived the same bypass rule for shear flows of an inviscid fluid, but based not on the principle of causality (meaning that the perturbation should vanish in the distant past) like Landau, but on the dissipativity principle (i.e., by adding a term with an infinitely small positive viscosity to the right-hand side of the Euler equation).

The problem of finding the analytical continuation of \({\cal D}(\omega)\) for equilibrium spherical stellar systems is much more complex than in a homogeneous plasma \cite{Lan_46}, in an infinite homogeneous gravitating medium \citep{PSB}, or in shear flows of fluid \citep{PS_22}.
Firstly, in an inhomogeneous medium, we deal with an infinite number of resonant denominators \(1/[\omega-\bn \cdot \bOm(\bJ)] \) rather than a single one \(1/(\omega-k v)\). 
Secondly, even in the simplest case, we deal with at least two-dimensional phase space in the action variables \(\bJ\), rather than one-dimensional, where we have to work with integrals containing only a single velocity component parallel to the fixed direction of the wave vector \(\mathbf{k}\). An exception was provided by \cite{BOY11}, where an artificial non-physical potential was used. This allowed the authors to reduce the problem to a one-dimensional one, albeit with a large number of resonance denominators of the form  \(1/[\omega-n\,\Omega(J)]\).

For spherical systems with a real gravitational potential, specifically in \cite{Kin_66} models, an attempt to construct an analytical continuation of the determinant ${\cal D}(\omega)$ to the lower half-plane was made by  \cite{Wei_94}. To do this, he approximated the function ${\cal D}(\omega)$ in the upper half-plane with a sum of rational functions, allowing for a straightforward analytical continuation to the lower half-plane. By obtaining an approximate expression for the analytical continuation of ${\cal D}(\omega)$, he found (for specific model parameters) the frequencies of weakly decaying Landau quasi-modes. While the results of this work are widely cited in the literature, from our perspective, they are not sufficiently convincing.

Another way to detect Landau exponential decay is to directly solve the evolutionary equation (or system of equations) for the Fourier harmonics of the perturbed DF $f_n(J;t)$. To do this, one needs to specify the initial DF $f(\bJ,0)$ and the corresponding perturbed potential $\Phi(\br,0)$. If the considered equilibrium state contains a Landau quasi-mode, it should manifest itself for \textit{any choice} of initial DF, since the determinant ${\cal D}(\omega)$ depends only on the properties of the unperturbed (background) system and is independent of the properties of the initial perturbation. Full consistency of the asymptotic behavior of the amplitude ${\hat\rho}_k(t)$ of the density perturbation $\rho(x,t)={\hat\rho}_k(t)\,e^{ikx}$ with Landau decay with the frequency $\omega_{\rm L}$ obtained from the condition ${\cal D}(\omega)=0$ was found for an infinite homogeneous medium \citep{PSB}. As for the amplitude of the total vorticity across the channel $N_k(t)=\int dy\, {\hat\zeta}_k(y,t)$ in the problem of shear flows, Landau decay here turned out to be only an intermediate asymptotic, which was replaced by algebraic decay \citep{PS_22}. However, even in this case, Landau decay during the exponential stage had the `correct' frequency $\omega_{\rm L}$, determined from the dispersion equation ${\cal D}(\omega)=0$.

The results obtained using the methods described above strongly depend on the choice of codes and numerical parameters: phase space grids, the number of retained Fourier harmonics for variables $\bw$, as well as the number of retained basis functions. Therefore, having a test perturbation for code verification is highly desirable. One such test perturbation has long been known. It involves shifting the entire spherical system. If this shift occurs, for example, along the $z$-axis by a small distance $\xi$, the perturbations of density and potential that arise are $\rho(r,\theta)=-\xi \rho_0'(r) \cos\theta$, $\Phi(r,\theta)=-\xi  \Phi_0'(r) \cos \theta$. This is a dipole shift perturbation corresponding to the spherical harmonic $P_{l=1}(\cos\theta)=\cos\theta$, where $P_l(x)$ is Legendre's polynomial. It is evident that the eigenfrequency $\omega$ corresponding to this perturbation is zero. This test has been repeatedly used in code verification for stability studies~\citep[e.g.,][]{Tre_05, PS_15}.

In this work, we propose another simple test perturbation that allows for an exact solution. It is applicable to models with ergodic DFs containing a single length parameter $\ell$.

This exact solution also helps clarify another issue regarding the correct definition of perturbation energy. The perturbation energy, being a quadratic quantity in amplitude, at first glance cannot be computed in linear theory. However, it can be shown that the system of linearized Boltzmann equation and Poisson equation admits a quadratic integral of motion, which closely resembles the total perturbation energy. Strictly speaking, this does not coincide with the actual energy, as its calculation requires knowledge of second-order perturbations in DF, potential, and density. The proposed test perturbation allows for evaluation of perturbation energy to any order and enables a comparison of these two second-order `energies'.

In Section 2, we will introduce the concept of a test perturbation and illustrate it with several examples of self-consistent equilibrium distribution function (DF) models that include the mode under consideration in their spectrum. Section 3 will delve deeper into the idea of perturbation energy. This can be formulated within the framework of linear theory. We will use a scale-invariant (dilation) perturbation as an example to compare the rigorously computed energy, which takes into account second-order perturbations, with the established expression for perturbation energy derived from linear theory. Finally, Section 4 will discuss the findings from our study.

\section{The idea of the test perturbation and several examples\\of relevant models}

Let the spherical model be described by an equilibrium DF containing a single characteristic scale with respect to the radial variable \(r\). We will call it the scaling factor and denote it as \(\ell\). For such models, the unperturbed potential and density have the following form:
\be
\Phi_0(r,\ell)=\frac{MG}{\ell}\,\phi\left(\frac{r}{\ell}\right),\ \ \ \rho_0(r,\ell)=\frac{M}{\ell^3}\,\varrho\left(\frac{r}{\ell}\right),
\label{eq:test_pho_rho}
\ee
where $\phi(x)$ and $\varrho(x)$ are related by Poisson’s equation
\be
\frac{1}{x^2}\,\frac{d}{dx} \left[x^2\,\frac{d\phi(x)}{d x}\right]=4\pi\,\varrho(x).
\ee
The DF is given by:
\be
F_0(\ce,\ell)=\frac{1}{(MG\,\ell)^{3/2}}\,{\cal F}(\ce),\ \ 0\le\ce\le\Psi(0)\equiv -\phi(0),
\ee
where the minus  dimensionless energy \(\ce\) of the star should also be considered as a function of \(v\), \(r\), and the scaling factor \(\ell\):
\be
\ce=\ce(r,v;\ell)=-\frac{\ell}{MG}\,\left[\frac{1}{2}\,v^2+\Phi_0(r,\ell)\right]=
-\frac{\ell}{MG}\,\left[\frac{1}{2}\,v^2+\frac{MG}{\ell}\,\phi\left(\frac{r}{\ell}\right)\right].
\ee
Here, the dimensionless function \({\cal F}(\ce)\) is normalized such that \(\int {\cal F}\,d^3\mathbf{r}\,d^{\,3}\mathbf{v}=1\).

It is entirely clear that if we fix the total mass $M$ but change $\ell$, we obtain the same equilibrium model, but with a different scaling factor: $\ell\to \ell + \delta \ell$. This means that the eigenfrequency of the mode $\om$ corresponding to such stretching/shrinking is zero. This fact can serve as a test for various codes when studying perturbations dynamics in spherical systems.

{ Let’s consider a few models of this type.}

\begin{itemize}
\item{\bf Isochrone model} \citep{Hen_60}

The model is characterised by the potential
\be
    \phi(x)=-\frac{1}{1+a}; \ \ a=\sqrt{1+x^2},
\ee
density
\be
    \varrho(x)=\frac{1}{4\pi}\,\frac{1+2\,a}{(1+a)^2\,a^3},
\ee
and DF
\begin{multline}
{\cal F}_{\rm H\acute{e}non}(\ce)= \frac{1}{\sqrt{2}\,(2\pi)^3}\,\frac{\sqrt{\ce}}{[2\,(1-\ce)]^4} \times \\
\times \Biggl[64\ce^4-240\ce^3+320\ce^2-66\ce+27
+3\,(16\ce^2+28\ce-9)\,\frac{\arcsin\sqrt{\ce}}{\sqrt{\ce\,(1-\ce)}}\Biggr];\ \ \ 0\le\ce\le \frac{1}{2}.
\label{eq:iso}
\end{multline}

\item {\bf Hernquist's model} \citep{Her_90} 

The model is characterised by the potential
\be
\phi(x)=-\frac{1}{1+x},
\ee
  density
  \be
\varrho(x)=\frac{1}{2\pi}\,\frac{1}{x\,(1+x)^3},
\ee
and DF
\be
{\cal F}_{\rm Hernquist}(\ce)=\frac{1}{\sqrt{2}\,(2\pi)^3}\,\frac{\sqrt{\ce}}{ (1-\ce)^2}\,
\left[(1-2\ce)\,(8\ce^2-8\ce-3)+\frac{3\,\arcsin\sqrt{\ce}}{\sqrt{\ce\,( 1-\ce)}}\right];\ \ 0\le \ce\le 1.
\ee

\item {\bf Jaffe model} \citep{Jaf_83} 

The model is characterised by the potential
\be
\phi(x)=-\ln\Bigl(1+\frac{1}{x}\Bigr),
\ee
  density
\be
\varrho(x)=\frac{1}{4\pi}\,\frac{1}{x^2\,(1+x)^2},
\ee
and DF
\be
{\cal F}_{\rm Jaffe}(\ce)=\frac{1}{2\pi^3}\,\Bigl[F_-(\sqrt{2\ce})-\sqrt{2} \,F_-(\sqrt{\ce})
-\sqrt{2}\,F_+(\sqrt{\ce})+F_+(\sqrt{2\ce})\Bigr];\ \ \ 0\le\ce <\infty,
\ee
where
$F_{\pm}(x)=e^{\mp x^2}\int_0^x dy\,e^{\pm y^2}$.

\item {\bf Plummer model} \citep{Plu_1911} 

The model is characterised by the potential
\be
\phi(x)=-\frac{1}{\sqrt{1+x^2}},
\ee
  density
\be
\varrho(x)=\frac{3}{4\pi}\,\frac{1}{(1+x^2)^{5/2}},
\ee
DF
\be
{\cal F}_{\rm Plummer}(\ce)=A\,\ce^{7/2}, \ \ A=\frac{3}{7}\,\frac{2^7}{ \sqrt{2}\,(2\pi)^3},
  \ \ \ \ 0\le \ce\le \Psi(0)=1.
\ee

\item  {\bf Polytropes}

The Plummer model is a special case of a series of polytropic models with the DF
\be
{\cal F}_{\rm polytropes}(\ce)=A_n\,\ce^{n-3/2}
\ee
and density
\be
\varrho(x)=\Lambda_n\,A_n\,\Psi^n(x), \quad \Lambda_n=\frac{1}{n!}\,(2\pi)^{3/2}\, \Gamma(n-\case{1}{2}),\ \ n>\frac{1}{2},
\ee
corresponding to $n=5$. 
For these models with arbitrary $n$, there isn't an explicit analytical expression for the potential $\phi(x)\equiv -\Psi(x)$. Instead, a corresponding nonlinear second-order equation for the potential arises from the Poisson equation (Lane-Emden equation). Research \citepalias[see][]{BT2008} indicates that polytropic models with $n>5$ possess infinite mass and are therefore irrelevant. Conversely, models with $1/2<n<5$ have a finite radius 
and must contain a scale-invariant mode in the spectrum, as this radius is the model's only length scale. Notably, for $n=1$, the Lane-Emden equation becomes linear and yields an analytical solution with finite radius $\ell=1$ and mass:
\be
\phi(x)=-\frac{\sin(\pi x)}{\pi\,x}, \ \ \varrho(x)=\frac{\sin(\pi x)}{4\,x },\ \ \ x\le 1,
\ee
\be
{\cal F}(\ce)=\frac{\sqrt{2}}{16\pi}\,\ce^{-1/2},\ \ 0\le \ce\le 1.
\ee
Let us note that models with $n<3/2$ have a positive sign of the energy derivative $E=-(M G/\ell)\,\ce$, that is, ${\cal F}'(\ce) <0$, and, in principle, may turn out to be unstable. We will not discuss this issue in more detail here.
\end{itemize}

Let’s consider the change in model parameters associated with the variation of the scale factor $\ell$, $\ell=\ell_0+\delta \ell$:
\be
\Phi(r,\ell)=\Phi_0(r,\ell_0)+\varepsilon\,\Phi_1(r,\ell_0)+\varepsilon^2\,\Phi_2(r,\ell_0)+{\cal O}(\varepsilon^3),
\ee
\be
\rho(r,\ell)=\rho_0(r,\ell_0)+\varepsilon\,\rho_1(r,\ell_0)+\varepsilon^2\,\rho_2(r,\ell_0)+{\cal O}(\varepsilon^3),
\ee
\be
F(\ce,\ell)=F_0(\ce,\ell_0)+\varepsilon\,f_1(\ce,\ell_0)+\varepsilon^2\,f_2(\ce,\ell_0)+{\cal O}(\varepsilon^3).
\ee
In this context, $\varepsilon={\delta \ell}/{\ell_0}\ll 1$ represents a small expansion parameter. We've expanded all quantities to the second order. While second-order quantities aren't necessary in linear theory, we include them to derive accurate expressions for potential and kinetic energy. These energies, being second-order quantities in terms of the perturbation amplitude $\varepsilon$, can't be computed merely as a bilinear form from first-order quantities. With the assumptions $G=M=\ell_0=1$, we can express the potential as follows:
\be
 \Phi_1=-(x\,\phi)',
 \label{eq:Phi1}
\ee
\be
  \Phi_2= {\case{1}{2}}\,\bigl[\,2\,(x\,\phi)'+x\,(x\,\phi)''\bigr],
 \label{eq:Phi2}
\ee
for density, we have 
\be
\rho_1=-\bigl(\,3\,\varrho+ x\,\varrho'),
\label{eq:rho1}
\ee
\be
\rho_2=
6\,\varrho+ 4\,x\,\varrho'+{\case{1}{2}}\, x^2\,\varrho'',
\label{eq:rho2}
\ee
and for the DF 
\be
 f_1= -{\case{3}{2}}\,{\cal F}+{\cal F}'(\ce)\,\bigl[\ce+(x\,\phi)'\bigr],
 \label{eq:f1}
\ee
\be
f_2= {\case{1}{2}}\left\{ {\case{15}{4}}\,{\cal F}-{\cal F}'\,\bigl[\,3\,\ce+3\,(x\phi)'+x\,(x\phi)''\bigr]+{\cal F}''\bigl[\ce+(x\phi)'\bigr]^2\right\}.
 \label{eq:f2}
\ee
The prime on functions denotes the derivative with respect to the corresponding argument. Using (\ref{eq:Phi1}), we write
\be
 f_1=-{\case{3}{2}}\,{\cal F}(\ce)+{\cal F}'(\ce)\,\bigl[\ce-\Phi_1(x)\bigr].
 \label{eq:test_F_1_Phi1}
\ee
It can be readily demonstrated that this type of perturbation results in zero mass perturbation in both the first and second orders: $\int \rho_{1,2}\,d^3\br=\int d^3\br \int f_{1,2}\,d^3\bv=0$.

The first-order perturbations $\Phi_1$, $\rho_1$, and $F_1$ constitute a test perturbation, which corresponds to an eigenfrequency $\omega=0$ in the eigenvalue problem. However, if we explore the dynamics of these perturbations using the solution to the system of evolutionary equations for the amplitudes of Fourier harmonics of the perturbed DF $f_n$, and if we define the initial DF as per (\ref{eq:f1}) and the potential as per (\ref{eq:Phi1}), we should find that $\partial f_n/\partial t=0$.

Indeed, the linearized kinetic equation for radial perturbations, where $f$ is equivalent to $F_1$ and $\Phi$ is equivalent to $\Phi_1$, in the context of action-angle variables, can be expressed as follows:
\be
\frac{\p f}{\p t}=-\Omega\,\frac{\p}{\p w}\,\Bigl(f+{\cal F}'\Phi\Bigr),
\label{eq:kin_main_1}
\ee
where $\Omega\equiv\Omega_R(\ce,L)$ is the frequency corresponding to radial action $J_R$, $\Omega_R=\partial H_0/\partial J_R$, $L=J_\theta+|J_\phi|$ is the angular momentum, and $w\equiv w_R$ is the angular variable conjugate to radial action, with $dw/dt=\Omega$.

In harmonics, we have:
\be
\frac{\p\,f_n}{\p t}=-\i\,n\Omega\,\Bigl(f_n+{\cal F}'\Phi_n\Bigr),
\label{eq:om_f_n}
\ee
where:
\[
f_n(\ce,L;t)=\oint dw\, f(\ce,L,w,t)\,e^{-\i nw}, \quad \Phi_n(\ce,L;t)=\oint dw\, \Phi(\ce,L,w,t)\,e^{-\i nw}
\]
and $\Phi(\ce,L,w,t)\equiv\Phi\Bigl(r(\ce,L,w),t\Bigr)$. From (\ref{eq:test_F_1_Phi1}), we have:
 \be
 f_n(\ce,L)=-{\cal F}'(\ce)\,\Phi_n(\ce,L), \ \ \ n\ne 0,
 \label{eq:test_dyn_eq}
 \ee
for non-zero harmonics. Using (\ref{eq:om_f_n}), we confirm that indeed $\partial f_n/\partial t=0$ for all $n\ne 0$, as it should be. It is also evident from (\ref{eq:om_f_n}) that the zeroth harmonic $f_{n=0}$ also remains constant.


The test of the evolutionary equation for the harmonics (\ref{eq:om_f_n}) was indeed carried out on the isochrone model (\ref{eq:iso}), for which it is relatively simple to obtain analytical expressions relating the radial coordinate $r$ to the action-angle variables, or equivalently, to the variables $\ce$, $L$, and the radial angular variable $w$.\footnote{The isochrone model offers further benefits, including the presence of explicit analytical expressions that connect the Hamiltonian $H_0(\bJ)=E$ to the action variables $\bJ=(J_R,J_\theta,J_\phi)$ \citepalias[see][eq. 3.226]{BT2008}. Another advantage of the isochrone potential is that  the radial frequency $\Omega(\bJ)$ depends only on the energy, $\Omega=[-2 E(\bJ)]^{3/2}=(2\ce)^{3/2}$ ($M=G=\ell=1$).}
Knowing the parametric relation between $r$ and $w$:
\be
 r(\ce,L,\xi)=\sqrt{\left(\frac{1-p\,\cos\xi}{2\ce}\right)^2-1},\ \ w=\xi-p\,\sin\xi, \ \ \ p=\sqrt{(1-2\ce)^2-2\ce\,L^2},
\ee
with $-\pi\leq\xi\leq\pi$ and $-\pi\leq w\leq \pi$, one can numerically perform the Fourier expansion of the radial angular variable $w$. By setting functions (\ref{eq:f1}) and (\ref{eq:Phi1}) as the initial perturbation of the DF $f_1$ and potential $\Phi_1$ respectively, and expanding them in harmonics:
$f_1(\ce,L,w)=(2\pi)^{-1}\sum f_n(\ce,L;0)\, e^{\i nw}$, $\Phi_1(\ce,L;w)=(2\pi)^{-1}\sum \Phi_n(\ce,L;0)\,e^{\i nw}$, we indeed obtain that $f_n(\ce,L;t)=f_n(\ce,L;0)$ for all $n$.

Furthermore, for this model, a check for the conservation of total mass is performed, i.e., the vanishing of the integral of the zeroth harmonic of the perturbed DF over the allowed region of the phase space $\bvpi=(\ce,L)$ of the model, 
\[
M_1 =(2\pi)^2\,\int\limits_0^{1/2} \dfrac{d\ce}{\Omega(\ce)} \int\limits_0^{L_{\rm circ}^2(\ce)}\, d(L^2)\Bigl\{2\pi\,\Bigl[-{\case{3}{2}}\,{\cal F}(\ce)+\ce\,{\cal F}'(\ce)\Bigr] -{\cal F}'(\ce)\,\Phi_{n=0}(\ce,L)\Bigr\}=0,
\]
where $L_{\rm circ}(\ce)=\dfrac{1-2\ce}{\sqrt{2\ce}}$ is the circular orbit line, and
$
\Phi_{n=0}(\ce,L)=2\pi\,\dfrac{(2\ce)^{3/2}}{\sqrt{4+L^2}}
$.

\section{Energy and Pseudoenergy of Perturbations}
\label{sec:energy}

In the monograph by \citetalias[][Section 5.4.2]{BT2008}, it is demonstrated within the linear approximation framework that for perturbations in systems with a decreasing ergodic distribution function (DF), $F'(E)<0$, and in the absence of external forces, there exists a quadratic integral of motion. This integral, known as the perturbation energy, is constructed solely from first-order quantities. We'll denote this as $E_{\rm BT}$:
\be
E_{\rm BT}=\varepsilon^2\,({K}+{P}),
\label{eq:E_BT}
\ee
where
\be
K=\frac{1}{2} \int\frac{f_1^2\,d^3\br\,d^3\bv}{-dF_0/dE},
\label{eq:K}
\ee
\be
P=\frac{1}{2} \int d^3\br\,\Phi_1(\br)\,\rho_1(\br)=
-\frac{1}{8\pi} \int d^3\br\,[\nabla \Phi_1(\br)]^2.
\label{eq:P}
\ee
In (\ref{eq:P}), it is taken into account that $\Delta \Phi=4\pi\rho$. Indeed, the expression for ${E}_{\rm BT}$ can be derived based on the concept proposed by \cite{NT99}. They suggested evaluating the work performed on the system by an external force $-\varepsilon\,\nabla \Phi_{\rm ext}$, which is treated as a first-order quantity. Consequently, we obtain:
\be
\frac{d{E}_{\rm BT}}{dt}=-\varepsilon^2 \int d^3\br \,d^3\bv\, f_1(\br,\bv,t)\,\bv\cdot\nabla\Phi_{\rm ext}.
\label{eq:dE_BT_dt}
\ee
Therefore, the quantity $E_{\rm BT}$ is commonly associated with the total energy of the perturbation. Sometimes $\varepsilon^2\,{K}$ is associated with the kinetic part, and $\varepsilon^2\,{P}$ with the potential part of it \citep[e.g.,][]{LB2021a}.\footnote{Recently, \cite{LB2021b} successfully extended the expression for perturbation energy to accommodate arbitrary non-ergodic systems with an integrable Hamiltonian. This generalization is particularly applicable to anisotropic spherical systems where $F=F(E,L)$.}

We aim to ascertain the validity of treating $E_{\rm BT}$ as the energy of a second-order perturbation. This is because it's derived within the confines of linear theory, without considering the contributions from second-order quantities. The scale-invariant mode, which allows us to explicitly derive perturbations for any order in perturbation theory, facilitates a comparison between the known accurate expression for second-order perturbation energy, denoted as $E_{\rm true}$, and the expression $E_{\rm BT}$.

For $E_{\rm true}$, taking into account the contributions of the second order, we have:
\be
 E_{\rm true}=\varepsilon^2 (K_{\rm true}+P_{\rm true}),
 \label{eq:E_true}
\ee
where
\be
 K_{\rm true}=\varepsilon^2\int d^3\br\, d^3\bv\, f_2(\br,\bv)\,\frac{v^2}{2}
 \label{eq:K_true}
\ee
is the second-order perturbation of kinetic energy, and the perturbation of potential energy can be expressed in two equivalent forms:
\be
 P_{\rm true}=-\frac{1}{8\pi\,G}\,\int d^3\br\left\{\,[\nabla\Phi_1(\br)]^2+2\nabla\Phi_0(\br)\,\nabla\Phi_2(\br)\right\},
 \label{eq:P_via_Phi'}
\ee
\be
 P_{\rm true}
  =-\frac{1}{8\pi\,G}\,\int d^3\br \,[\nabla\Phi_1(\br)]^2
  +\int d^3\br\,\Phi_0(\br)\int d^3\bv\, f_2\,.
  \label{eq:P_via_rho}
\ee
Combining the contributions containing $f_2$ in (\ref{eq:E_true}), we can write:
\be
 E_{\rm true}=\int d^3\br\, d^3\bv\, f_2(\br,\bv)\,E +P.
\label{eq:E_via_f}
\ee
Here, $P$ is a bilinear form defined in (\ref{eq:P}), and $E=\frac{1}{2}\,v^2+\Phi_0(\mathbf{r})$ is the energy of a star in the unperturbed potential $\Phi_0(\mathbf{r})$.

First, we need to confirm that the accurate second-order potential energy, denoted as \(P_{\rm true}\), can only be correctly derived when we consider second-order perturbations. This can be done using either equation (\ref{eq:P_via_Phi'}) or equation (\ref{eq:P_via_rho}). Let's begin with the precise expression for the total potential energy. For models with a single scaling factor, this expression is as follows:
\be
P_{\rm total}=\frac{G M^2}{\ell}\,V,\ \ V=-\frac{1}{2}\,\int x^2\,\bigl[\phi'(x)\bigr]^2\,dx.
\label{eq:P_total}
\ee
Assuming \(M=G=1\), \(\ell=1+\varepsilon\), and expanding in \(\varepsilon\), we obtain in the second order:
\be
P_{\rm true}=V.
\label{eq:P_equal_V}
\ee
On the other hand, from (\ref{eq:P_via_Phi'}), we have:
\be
P_{\rm true}=-\frac{1}{2}\int x^2 \bigl[\Phi_1'^2(x)+2 \Phi_0'(x)\,\Phi_2'(x)\bigr]\,dx.
\label{eq:Phi0_Phi2}
\ee
Taking into account that according to (\ref{eq:Phi1}) and (\ref{eq:Phi2}),
\(\Phi_0'=\phi'\), \(\Phi_1'=-(x\phi)''\), \(\Phi_2'=\frac{1}{2}\,x^{-2}\,\bigl[x^3 (x\phi'')\bigr]'\),
and after a chain of integration by parts, we obtain:
\be
P_{\rm true}=
-\frac{1}{2}\,\int x^2\,\phi'^2\,dx=V,
\label{eq:P_V}
\ee
which coincides with (\ref{eq:P_equal_V}). This underscores the importance of incorporating second-order perturbations for an accurate computation of potential energy.

Next, we calculate the kinetic energy using equations (\ref{eq:K_true}) and (\ref{eq:f2}). After some complex calculations (detailed in Appendix A), we find:
\be
K_{\rm true}=-{\case{1}{2}}\,V,
\label{eq:virial}
\ee
as it should be. Given that the system maintains equilibrium, the virial relation should be valid across all orders of perturbation theory. Specifically, for the second order, we should have \(2K_{\rm true}+P_{\rm true}=0\). Indeed, we observe that the accurate expression for the perturbation energy, denoted as \(E_{\rm true}\), is provided by equation (\ref{eq:E_via_f}), which does not coincide with the formula \(E_{\rm BT}\) (\ref{eq:E_BT}), as \(E_{\rm BT}=P+K\ne P_{\rm true}+K_{\rm true}=E_{\rm true}\). Nevertheless, it turns out that their time derivatives are equal. Let's demonstrate this.

In the second order, from the Boltzmann equation we have:
\be
\left(\frac{\p }{\p t}+\bv\,\nabla-\nabla\Phi_0\,\frac{\p }{\p \bv}\right)\,f_2=
\nabla \Phi_2\,\frac{\p F_0}{\p \bv}+\nabla\,\Phi_1\,\frac{\p f_1}{\p \bv}+\nabla\Phi_{\rm ext}\,\frac{\p f_1}{\p\bv}.
\label{kin_eq_E}
\ee
Here we’ve introduced an external potential, \(\Phi_{\rm ext}(\mathbf{r})\). This potential is responsible for changes in the total energy of the system. The gravitational force it generates, \(-\varepsilon\nabla \Phi_{\rm ext}\), performs work on stars of the system.

Multiplying both sides of (\ref{kin_eq_E}) by \(E\), integrating over the phase space, and taking into account that \(E\) is an integral of unperturbed motion, we get:
\be
\frac{d}{dt} \int d^3\br\, d^3\bv\, f_2(\br,\bv)\,E=\int d^3\br\, d^3\bv \,E\,\left(\nabla \Phi_2\,\frac{\p F_0}{\p \bv}+\nabla\,\Phi_1\,\frac{\p f_1}{\p \bv}+\nabla\Phi_{\rm ext}\,\frac{\p f_1}{\p\bv}\right).
\label{eq:d_dt}
\ee
The first term on the right-hand side of (\ref{eq:d_dt}) becomes zero due to the antisymmetry of the integrand with respect to \(\mathbf{v}\), since \(E\,\partial F_0/\partial \mathbf{v} = E\,\mathbf{v}\,F_0'(E)\). The second term, after a series of transformations, turns into \(-dP/dt\). Indeed, we have for it: 
\begin{multline}
\int d^3\br\, \nabla\,\Phi_1 \int d^3\bv \,E\,\frac{\p f_1}{\p \bv}=-\int d^3\br\, \nabla \Phi_1 \int d^3\bv\, (\bv\,f_1)\\=\int d^3\br\, \Phi_1(\br)\,\nabla \int d^3\bv\,(\bv\,f_1)=-\int d^3\br\, \Phi_1\,\frac{\p \rho_1}{\p t}=-\int d^3\br\, \frac{\p\Phi_1}{\p t}\,\rho_1(\br)\\=-\frac{1}{2}\int d^3\br\,\Bigl(\Phi_1\,\frac{\p \rho_1}{\p t}+\frac{\p\Phi_1}{\p t}\,\rho_1\Bigr)=-\frac{d}{dt}\,\Bigl(\frac{1}{2}\,\int d^3\br \,\rho_1\,\Phi_1\Bigr)=-\frac{dP}{d t}.
\label{eq:dot_Phi_rho}
\end{multline}
The third term transforms into \(-\int d^3\mathbf{r}\,d^3\mathbf{v}\,f_1\,\mathbf{v}\cdot\nabla\Phi_{\rm ext}\):
\be
\int d^3 \br\, \nabla\Phi_{\rm ext}\int d^3 \bv\,E\,\frac{\p f_1}{\p\bv}=
-\int d^3\br\,d^3\bv\,f_1\,\bv\!\cdot\!\nabla\Phi_{\rm ext}.
\label{eq:Phi_ext}
\ee
Finally, from (\ref{eq:d_dt}), (\ref{eq:dot_Phi_rho}), and (\ref{eq:Phi_ext}), we obtain:
\[
\varepsilon^2\,\frac{d }{d t}\,\Bigl(\int d^3\br\, d^3\bv\, f_2\,\frac{v^2}{2}+
 P \Bigr)=-\varepsilon^2\int d^3\br\,d^3\bv\,f_1\,\bv\!\cdot\!\nabla\Phi_{\rm ext},
\]
or, taking into account (\ref{eq:E_via_f}):
\be
\frac{d E_{\rm true}}{dt}=-\varepsilon^2\int d^3\br\,d^3\bv\,f_1\,\bv\!\cdot\!\nabla\Phi_{\rm ext}.
\label{eq:dE_true_dt}
\ee
Comparing the right-hand sides of (\ref{eq:dE_true_dt}) and (\ref{eq:dE_BT_dt}), we find that \(dE_{\rm BT}/dt=dE_{\rm true}/dt\). This means that the actual total energy of the perturbations, \(E_{\rm true}\), is different from the bilinear construction often referred to as the perturbation energy, \(E_{\rm BT}\), by a constant value. This suggests that both of these second-order quantities are conserved in the absence of external forces during the evolution of the system. Hence, \(E_{\rm BT}\), which is defined by the relationships (\ref{eq:E_BT}), (\ref{eq:K}), and (\ref{eq:P}), and is similar to the linear theory of fluid shear flows, can be appropriately termed as a `pseudoenergy'. It's worth noting that in shear flow theory, there's also a concept of a pseudoenergy integral, which is constructed as a bilinear form of first-order perturbations. Pseudoenergy is different from the true energy, which should be calculated considering second-order perturbations~\citep[see][]{Hel_85,PS_22}.
In particular, for the isochrone model, we calculated the pseudoenergy of the scale-invariant  mode (see Appendix B):
\be
E_{\rm BT}=-0.0209\,\varepsilon^2,
\label{eq:E_BT_negative}
\ee
while the true energy is
\be
E_{\rm true}= -0.0594\,\varepsilon^2,
\ee
that is, the difference is approximately threefold. It’s important to note that the pseudoenergy and the true energy can have different signs. As proven by \cite{LB2021a}, the pseudoenergy \(E_{\rm BT}\) of the eigenmodes of systems with a decreasing ergodic DF, \(F_0'(E)<0\) (i.e., the van Kampen mode, as there are no other eigenmodes in such systems, except for the dilation model), is strictly positive.\footnote{It may appear that this contradicts the earlier finding that the pseudoenergy of the scale-invariant mode is negative (as seen in (\ref{eq:E_BT_negative})). However, there’s actually no contradiction. The proof of pseudoenergy’s positivity by \cite{LB2021a} depends on the condition \(\omega^2>0\). If \(\omega^2=0\), the sign of the pseudoenergy isn’t uniquely determined.}
The true energy can indeed have any sign. This fact could be crucial for the concept of constructing the thermodynamics of star clusters based on the excitation of van Kampen waves, as suggested by \cite{LB2021a}. The positive sign of energy might be a key factor for successfully implementing this idea. However, for our scale-invariant perturbation, the true energy is actually \textit{negative}. This is evident from the virial relation, which states that the energy of perturbations is equal to half of the potential energy, \(E_{\rm true}=\varepsilon^2 (K_{\rm true}+P_{\rm true})=\frac{1}{2}\,\varepsilon^2 P_{\rm true}\). As seen from (\ref{eq:P_V}), this value is negative, \(P_{\rm true}<0\).

\section{Conclusions}

The paper emphasizes the importance of a specific radial stationary perturbation that can be explicitly expressed in terms of distribution function, density, and gravitational potential. This perturbation is relevant for ergodic systems with a \textit{single} length-scale factor.

It is highlighted that when solving the initial problem for the perturbation of the DF \(f(\mathbf{r},\mathbf{v};t)\), it should confirm phase-space distribution conservation, \(f(\mathbf{r},\mathbf{v};t)=f(\mathbf{r},\mathbf{v};0)\), if the latter is chosen as the initial condition. 
When  the solving for eigenvalues using the conventional matrix method~\citep[e.g.][]{PS81} this perturbation should yield an eigenfrequency \(\omega=0\) along with the known eigenfunctions. The isochrone model was indeed subjected to this test. Surprisingly, the test proved to be quite complex and even paradoxical. The procedure followed will be elaborated upon in a separate study. 

The concept of perturbation energy in the linear perturbation theory in collisionless stellar systems is also analyzed. While the true perturbation energy, being a second-order quantity in the perturbation amplitude, cannot be computed within the linear theory, a bilinear form of first-order quantities can be constructed. This form closely resembles energy and represents the sum of two contributions sometimes referred to as `kinetic' and `potential' perturbation energies.

Using a test perturbation for which expressions in any order of the perturbation theory can be obtained, it's demonstrated that the expressions for `kinetic' and `potential' energy expressions derived from linear theory do not align with the correct expressions for kinetic and potential energies, which consider second-order perturbations. The paper concludes by highlighting an important observation: the integral of motion, represented by the correct expression for perturbation energy, and the integral corresponding to the `energy’ constructed within linear theory (referred to as pseudoenergy) do not match. However, their difference is a time-independent constant. Interestingly, these quantities can exhibit opposite signs. This discrepancy could have significant implications for problems related to the application of van Kampen modes to stellar systems.

\bigskip

\section*{Funding}

This work was supported by the Foundation for the Advancement of Theoretical Physics and Mathematics ``BASIS'' (grant No. 20-1-2-33), the Program of the Presidium of the Russian Academy of Sciences No. 28 "Space: Research of Fundamental Processes and Their Interrelations" (subprogram II ``Astrophysical Objects as Space Laboratories''), as well as the Ministry of Science and Higher Education of the Russian Federation (Ilia Shukhman).

\bibliographystyle{aa}
\bibliography{main}   

\appendix

\section{Calculation of True Kinetic Energy for the Scale-invariant Mode}
\setcounter{equation}{0}
\renewcommand{\theequation}{\Alph{section}.\arabic{equation}}

Let's demonstrate that the kinetic energy of the perturbation \(K_{\rm true}\), calculated using the second-order perturbed DF \(f_2\) (Eq.~\ref{eq:f2}), satisfies the virial relation \(K_{\rm true} = -\frac{1}{2} P_{\rm true}\), or equivalently, \(K_{\rm true} = -\frac{1}{2} V\), where \(P_{\rm true} = V = -\frac{1}{2}\int dx\,x^2\,\phi'^2\) is the potential energy.

Starting with:
\be
K_{\rm true}={\case{1}{2}}\int d^3\br\,d^3\bv f_2 v^2,
\ee
or
\be
K_{\rm true}={\case{1}{4}}\int d^3x\int d\ce\, 4\pi\,(2\Psi-2\ce)^{3/2} \big\{{\case{15}{4}}\,{\cal F}-{\cal F}'[3\ce+3(x\phi)'+x\,(x\phi)'']+{\cal F}''[\ce+(x\phi)']^2\big\}.
\label{eq:Ktrue}
\ee
we can break down \(K_{\rm true}\) into six terms: \(K_{\rm true}=\sum_{s=1}^6 K^{(s)}\), where:
\be
K^{(1)}={\case{15}{16}}\,\int d^3x \int d\ce\, 4\pi\,(2\Psi-2\ce)^{3/2}\,{\cal F},
\nonumber
\ee
\be
K^{(2)}=-{\case{3}{4}}\,\int d^3x \int d\ce\, 4\pi\,(2\Psi-2\ce)^{3/2}\,{\cal F}'\ce,
\nonumber
\ee
\be
K^{(3)}=-{\case{1}{4}}\,\int d^3x\,[3(x\phi)'+x\,(x\phi)''] \int d\ce\, 4\pi\,(2\Psi-2\ce)^{3/2}\,{\cal F}',
\nonumber
\ee
\be
K^{(4)}={\case{1}{4}}\,\int d^3x \int d\ce\, 4\pi\,(2\Psi-2\ce)^{3/2}\,{\cal F}''\ce^2,
\nonumber
\ee
\be
K^{(5)}={\case{1}{2}}\,\int d^3x\,(x\phi)' \int d\ce\, 4\pi\,(2\Psi-2\ce)^{3/2}\,{\cal F}''\ce,
\nonumber
\ee
\be
K^{(6)}={\case{1}{4}}\,\int d^3x\,[(x\phi)']^2 \int d\ce\, 4\pi\,(2\Psi-2\ce)^{3/2}\,{\cal F}''.
\nonumber
\ee

\bigskip

\centerline{(i). \textit{Calculation of} \(K^{(1)}\)}
\begin{multline}
K^{(1)}={\case{15}{8}}\,\int d^3x \int d\ce\, 4\pi\,(2\Psi-2\ce)^{1/2}\,{\cal F}
\,(\Psi-\ce) = \\={\case{15}{8}}\,\int d^3x\,(\varrho\,\Psi)-{\case{15}{8}}\int d^3x \int d\ce\, 4\pi\,(2\Psi-2\ce)^{1/2}\,{\cal F}\,\ce={\case{15}{8}}\,[-2V-(-{\case{3}{2}})\,V]=-{\case{15}{16}}\,V.
\nonumber
\end{multline}
Here, we used the relation:
\begin{multline}
 4\pi \int d^3x\,\int d\ce\,
(2\Psi-2\ce)^{1/2}\,{\cal F}\,\ce = \\=-\int d^3x\,\int d^3u\, \Bigl({\case{1}{2}}\,u^2+\phi\Bigr)\,{\cal F}=
-K_0-2P_0={\case{1}{2}}\,V-2V=-{\case{3}{2}}\,V,
\label{eq:virial_0}
\end{multline}
where, according to the virial theorem, the kinetic energy \(K_0\) for the unperturbed state and its potential energy \(P_0\) are related by \(K_0=-\frac{1}{2}\,{P_0}=-\frac{1}{2}\,V\).

\medskip

\centerline{(ii). \textit{Calculation of} \(K^{(2)}\)}

\medskip

After integrating by parts in the integral over \(\ce\), we have:
\begin{multline}
K^{(2)}={\case{3}{4}}\,\int d^3x \int d\ce\, 4\pi\,(2\Psi-2\ce)^{1/2}\,{\cal F}
(2\Psi-5\ce) = \\={\case{3}{2}}\int d^3x\,(\varrho\,\Psi)-{\case{15}{4}}\int d^3x \int d\ce\,4\pi\,(2\Psi-2\ce)^{1/2}\,{\cal F}\,\ce={\case{3}{2}}\,(-2V)-{\case{15}{4}}\,(-{\case{3}{2}}\,V)={\case{21}{8}}\,V.
\nonumber
\end{multline}


\centerline{(iii). \textit{Calculation of} \(K^{(3)}\)}

\smallskip

Again, integrating by parts in the integral over \(\ce\), we find:
\be
K^{(3)}=-{\case{3}{4}}\,\int d^3x\,[3(x\phi)'+x\,(x\phi)'']\,\int d\ce\, 4\pi\,(2\Psi-2\ce)^{1/2}\,{\cal F}'=-{\case{3}{4}}\,\int d^3x\,\varrho\,[3(x\phi)'+x\,(x\phi)''].
\nonumber
\ee

\smallskip

\centerline{(iv). \textit{Calculation of} \(K^{(4)}\)}

\medskip

After two integrations by parts with respect to \(\ce\), we obtain:
\be
K^{(4)}={\case{85}{16}}\,V+{\case{3}{4}}\int d^3x \,\Psi^2\,\frac{d\varrho}{d\Psi},
\nonumber
\ee
where we have used the relation:
\be
\int d\ce\,4\pi\, \frac{\cal F}{(2\Psi-2\ce)^{1/2}}=\frac{d}{d\Psi} \int 4\pi\, d\ce\, {\cal F}\, (2\Psi-2\ce)^{1/2}=\frac{d\varrho}{d\Psi}.
\nonumber
\ee

\smallskip

\centerline{(v). \textit{Calculation of} \(K^{(5)}\)}

\medskip

Again, after repeated integration by parts with respect to \(\ce\), we have:
\be
K^{(5)}=-{\case{15}{4}}\int d^3x \,\varrho\,(x\phi)' +
{\case{3}{2}}\int d^3x \,\Psi\, (x\phi)' \,\frac{d\varrho}{d\Psi}.
\nonumber
\ee

\smallskip

\centerline{(vi). \textit{Calculation of} \(K^{(6)}\)}

\medskip

Finally, performing the double integration by parts, we obtain:
\be
K^{(6)}={\case{3}{4}}\int d^3x\,[(x\phi)']^2\,\frac{d\varrho}{d\Psi}.
\nonumber
\ee

Adding up all 6 contributions, one can have:
\begin{multline}
K_{\rm true}=\sum\limits_{s=1}^6 K^{(s)}=\Bigl[-{\case{15}{16}}\Bigr]+\Bigl[{\case{21}{8}}\Bigr]+
\Bigl[-{\case{3}{4}}\,\int d^3x\,\varrho\,[3(x\phi)'+x\,(x\phi)''\Bigr]\\+
\Bigl[{\case{85}{16}}\,V+{\case{3}{4}}\int d^3x \,\Psi^2\,\frac{d\varrho}{d\Psi}\Bigr]+
\Bigl[-{\case{15}{4}}\int d^3x \,\varrho\,(x\phi)' +
{\case{3}{2}}\int d^3x \,\Psi\, (x\phi)' \,\frac{d\varrho}{d\Psi}\Bigr]+
\Bigl[{\case{3}{4}}\int d^3x\,[(x\phi)']^2\,\frac{d\varrho}{d\Psi}\Bigr]\\
=7\,V+
\int d^3x\,\varrho\,\Bigl[-{\case{9}{4}}\,(x\phi)'-{\case{3}{4}}\,x\,(x\phi)''-
{\case{15}{4}}\,(x\phi)'\Bigr]
+\int d^3x \,\frac{d\varrho}{d\Psi}\,\Bigl[{\case{3}{4}}\,\Psi^2+{\case{3}{2}}\,\Psi\,(x\phi)'+
{\case{3}{4}}\,[(x\phi)']^2\Bigr].
\nonumber
\end{multline}
Taking into account that \(\Psi=-\phi\), we write:
\begin{multline}
K_{\rm true}=
7\,V+
\int d^3x\,\varrho\,\Bigl[-{\case{9}{4}}\,(x\phi)'-{\case{3}{4}}\,x\,(x\phi)''-
{\case{15}{4}}\,(x\phi)'\Bigr] + \\
+\int d^3x \,\frac{d\varrho}{d\phi}\,\Bigl[-{\case{3}{4}}\,\phi^2+{\case{3}{2}}\,\phi\,(x\phi)'-
{\case{3}{4}}\,[(x\phi)']^2\Bigr].
\nonumber
\end{multline}
We split \(K_{\rm true}=S_1+S_2+S_3\), where:
\be
S_1=7\,V,
\nonumber
\ee
\begin{multline}
S_2=\int d^3x\,\varrho\,\Bigl[-{\case{9}{4}}\,(x\phi)'-{\case{3}{4}}\,x\,(x\phi)''-
{\case{15}{4}}\,(x\phi)'\Bigr] = \\=\int dx\,x\,(x\,\phi''+2\phi') \Bigl(-{\case{15}{2}}\,x\phi'-
{\case{3}{4}}\,x^2\,\phi''-6\phi\Bigr)=-9V-{\case{3}{4}}\int dx\,x^4\,\phi''^2,
\nonumber
\end{multline}
\begin{multline}
S_3=\int d^3x \,\frac{\varrho'}{\phi'}\,\Bigl[-{\case{3}{4}}\,\phi^2+{\case{3}{2}}\,\phi\,(x\phi)'-
{\case{3}{4}}\,[(x\phi)']^2\Bigr]=-{\case{3}{4}}\int d^3x\,x^2\varrho'\phi' = \\=
-{\case{3}{4}}\int dx\,x^4 \Bigl[\phi'''-\frac{2}{x^2}\,\phi'+\frac{2}{x}\,\phi''\Bigr]\,\phi'=
{\case{3}{2}}\,V+{\case{3}{4}}\int dx\,x^4\,\phi''^2,
\nonumber
\end{multline}
where we used the relations \(d\varrho/d\phi=\rho'/\phi'\) and \(\varrho=(\phi''+2\phi'/x)/(4\pi)\). Finally, summing up all three contributions \(S_{1,2,3}\), we obtain the required relation:
\be
K_{\rm true}=\bigl(7-9+{\case{3}{2}}\bigr)\,V=-{\case{1}{2}}\,V.
\label{eq:K_true-app}
\ee

\section{Pseudoenergy $E_{\rm BT}$ of Scale-invariant Mode for the Isochrone Model}
\setcounter{equation}{0}
\renewcommand{\theequation}{\Alph{section}.\arabic{equation}}

Using (\ref{eq:K}), (\ref{eq:P}) and (\ref{eq:E_BT}), we write the expression for pseudoenergy as 
\be
  E_{\rm BT}=\frac{1}{2}\int d^3\br\int d^3\bv\,\Bigl(\frac{f_1^2}{{\cal F}'}+f_1\,\Phi_1\Bigr)=\frac{1}{2}\int d^3x\int d\ce\,
4\pi \,\sqrt{2\Psi-2\ce}\,\frac{f_1\,(f_1+{\cal F}'\Phi_1)}{{\cal F'}}.
\ee
Since according to (\ref{eq:test_F_1_Phi1}) and (\ref{eq:Phi1}) the perturbed DF $f_1$ and potential $\Phi_1$ are, respectively,
\be
f_1=-{\case{3}{2}}\,{\cal F}+\ce\,{\cal F}'-\Phi_1\,{\cal F}', \ \ \ \Phi_1=-(x\phi)',
\ee
and for isochrone model (\ref{eq:iso}) $\phi(x)\equiv\Phi_0(x)=-[1+\sqrt{1+x^2}]^{-1}$,  then
\begin{multline}
  E_{\rm BT}=\frac{1}{2}\int d^3x\int d\ce\,
4\pi \,\sqrt{2\Psi-2\ce}\,\frac{\Bigl[\Bigl(-{\case{3}{2}}\,{\cal F}+\ce\,{\cal F}'\Bigr)+(x\phi)'\,{\cal F}'\Bigr]\cdot
\Bigl(-{\case{3}{2}}\,{\cal F}+\ce\,{\cal F}'\Bigr)}{{\cal F'}}\\
=I+\frac{1}{2}\int d^3x\int d\ce\,
4\pi \,\sqrt{2\Psi-2\ce}\Bigl[\,\ce\,(-3{\cal F}+\ce\,{\cal F}')+(x\phi)'(-{\case{3}{2}}\,{\cal F}+\ce\,{\cal F}')\Bigr].
\label{eq:E_sum}
\end{multline}
Here $I$ is the integral 
\be
I=\frac{9}{8}\, (4\pi)^2\int\limits_0^\infty dx\,x^2\!\! \int\limits_0^\Psi d\ce\, \sqrt{2\Psi-2\ce}\,\frac{{\cal F}^2}{{\cal F}'}.
\label{eq:I}
\ee
Let's split the integral 
 in (\ref{eq:E_sum}) into two parts:
\be
E_{\rm BT}=I+E^{(1)}+E^{(2)},
\label{eq:E_sum1}
\ee
where
\be
E^{(1)}=\frac{1}{2}\int d^3x\int d\ce\,
4\pi \,\sqrt{2\Psi-2\ce}\bigl(-3\ce\,{\cal F}+\ce^2\,{\cal F}'\bigr),
\nonumber
\ee
\be
E^{(2)}=\frac{1}{2}\int d^3x\, (x\phi)' \int d\ce\,
4\pi \,\sqrt{2\Psi-2\ce}\,(-{\case{3}{2}}\,{\cal F}+\ce\,{\cal F}').
\nonumber
\ee
After transformations, similar to those performed in Appendix A, we get
\be
E^{(1)}={\case{37}{8}}\,V-{\case{1}{2}}\int d^3x\,\phi^2\,\frac{d\varrho}{d\phi},
\nonumber
\ee
\be
E^{(2)}=-{\case{3}{2}}\int d^3x\, (x\phi)'\,\varrho+
{\case{1}{2}}\int d^3x\,(x\phi)'\,\phi\,\frac{d\varrho}{d\phi}.
\nonumber
\ee
In total, $E^{(1)}+E^{(2)}$ is
\begin{multline}
E^{(1)}+E^{(2)}={\case{37}{8}}\,V-{\case{3}{2}}\int d^3x\, (x\phi)'\,\varrho+
{\case{1}{2}}\int d^3x\,\frac{\varrho'}{\phi'}\,[(x\phi)'\phi-\phi^2]
\\={\case{37}{8}}\,V-{\case{3}{2}}\int d^3x\, (x\phi)'\,\varrho+
{\case{1}{2}}\int d^3x\,x\phi\,\varrho'\\=
{\case{37}{8}}\,V-{\case{3}{2}}\int d^3x\, (x\phi'+\phi)\,\varrho
-{\case{1}{2}}\int d^3x\, \varrho\,(3\,\phi+x\phi')\\=
{\case{37}{8}}\,V-3\int d^3x\,(\varrho\,\phi)-2\int dx\, x^3\,\phi'\Bigl(\phi''+\frac{2}{x}\,\phi'\Bigr)\\={\case{37}{8}}\,V-3\int d^3x\,(\varrho\,\phi)+
3\int dx\,x^2\phi'^2 - 4\int dx\,x^2\phi'^2\\={\case{37}{8}}\,V-6V+2V={\case{5}{8}}\,V.\phantom{fffffffffffff}
\label{eq:E1+E2}
\end{multline}
For the isochrone model, where $\phi'=\dfrac{x}{a\,(1+a)^2}$, $a=\sqrt{1+x^2}$, one obtains:
\be
V=-{\case{1}{2}}\int\limits_0^\infty dx\,x^2\phi'^2=-{\case{1}{2}}\int\limits_1^\infty \frac{da}{a\,\sqrt{a^2-1}}\,\Bigl(\frac{a-1}{a+1}\Bigr)^2,
\nonumber
\ee
and after substitution $\dfrac{a-1}{a+1}=s^2$ 
\be
V=-\int\limits_0^1 \frac{s^4\,ds}{1+s^2}=-(\case{1}{4}\,\pi-\case{2}{3})=-0.1187.
\label{eq:V}
\ee
 Integral (\ref{eq:I}) for $I$ is calculated numerically: 
\be
I= 0.0533.
\label{eq:I1}
\ee
Finally, we find from (\ref{eq:E_sum1}) -- (\ref{eq:I1}) 
\be
 E_{\rm BT}=I+{\case{5}{8}}\,V=-0.0209.
\ee

\end{document}